\documentclass[11pt]{article}
\usepackage{type1cm,amsmath}
\usepackage[psamsfonts]{amssymb}
\usepackage{cite}
\usepackage{graphicx}

\def \foot {\footnote}

\begin{document}

\title{\textbf{Strings on type IIB pp-wave backgrounds with
interacting massive theories on the worldsheet}}
\author{Alin Tirziu and Paul Fendley\\
\\ Department of Physics, University of Virginia \\
Charlottesville, VA 22904-4714 \\
{\tt at2j@virginia.edu, fendley@virginia.edu}}

\maketitle

\begin{abstract}
We consider superstring theories on pp-wave backgrounds which
result in an integrable ${\cal N}=(2,2)$ supersymmetric
Landau-Ginzburg theory on the worldsheet. We obtain exact
eigenvalues of the light-cone gauge superstring hamiltonian in the
massive and interacting world-sheet theory with superpotential
$Z^3-Z$. We find the modes of the supergravity part of the string
spectrum, and their space-time interpretation. Because the system
is effectively at strong coupling on the worldsheet, these modes
are not in one-to-one correspondence with the usual type IIB
supergravity modes in the $p_{-} \to 0$ limit. However, the above
correspondence holds in the $\alpha'\to 0$ limit.
\end{abstract}

\section{Introduction}

Since the discovery of the AdS/CFT correspondence \cite{malda2},
superstring theories on Ramond-Ramond backgrounds have been
studied extensively. These theories are also interesting because
the RNS formalism cannot be applied; one needs to use the
Green-Schwarz formalism. The Penrose limit of the $AdS_{5}\times
S_{5}$ background gives a pp-wave background which has been
studied in the context of AdS/CFT correspondence \cite{beren}.
These maximally supersymmetric pp-wave solutions of IIB
supergravity were considered in \cite{blau}. The maximally
supersymmetric pp-wave background supported by a constant R-R
five-form field strength is also a useful framework, since the
string theory on it can be solved exactly \cite{metsaev1,
metsaev2}. Since D-branes are expected to be dual to
non-perturbative objects in gauge theory, they play an important
role in AdS/CFT correspondence. D-branes on maximally
supersymmetric pp-wave backgrounds have been studied in
\cite{bain1, dabho1, gaberd1, skenderis}.

One remarkable feature of the theories studied in \cite{metsaev1,
metsaev2} is that in the light-cone gauge GS formalism, they
result in massive theories on the world sheet. This means that one
cannot use conformal field theory when studying the physics from
the two-dimensional point-of-view. The world-sheet theories of
\cite{metsaev1, metsaev2} are free, however, so the spectrum of
the light-cone gauge Hamiltonian can still be easily found.
However, this is not true for another interesting class of pp-wave
type backgrounds which are solutions of type IIB supergravity with
non-constant R-R five-form field strength \cite{malda1}.  In the
case of flat transverse space these backgrounds provide exact
solutions of superstring theory \cite{berk}.  String theories on
these backgrounds give interacting ${\cal N}=(2,2)$ or ${\cal
N}=(1,1)$ supersymmetric field theories on the worldsheet. The
first class is parametrized in terms of an arbitrary holomorphic
function which becomes the worldsheet superpotential in the
light-cone gauge superstring action. Generic superpotentials are
massive, so conformal field theory cannot be applied. One can
obtain exact information about the spectrum in generic ${\cal
N}=(2,2)$ theories \cite{CFIV}, but except in simple cases it is
difficult to extract quantitative information from this formalism.

It is therefore useful to study pp-wave backgrounds which lead to
worldsheet superpotentials describing integrable massive theories.
In these theories, several techniques make possible the
computation of the spectrum as a function of worldsheet mass and
size. The technique we will utilize here is a generalization of
the thermodynamic Bethe ansatz \cite{yang} to compute
excited-state energies \cite{BLZ}. In fact, the finite-size
spectrum for the simplest ${\cal N}=(2,2)$ massive field theory,
that with superpotential $Z^3 - Z$, has already been derived
\cite{fendley4}.

In this paper, we discuss how to utilize these results in string
theory on type IIB supergravity backgrounds. The transverse target
space is taken flat and non-compact. Two of the eight transverse
space directions are interacting while the others are free. On
non-compact two dimensional space, the 2d quantum field theory
with the above superpotential has the spectrum consisting of kinks
and antikinks interpolating between the two supersymmetric vacua.
When considered on a cylinder, as necessary for closed strings,
the states consist of pairs of interacting kinks and antikinks
\cite{fendley4}. In target space they correspond to strings
interpolating between two planes sitting at the minima of the
superpotential.

The supergravity part of the string spectrum can be obtained by
considering all states contributing to the string's point particle
limit, which can be obtained by taking either the $\alpha'\to 0$
or $p_{-} \to 0$ limit. In the first case we obtain the type IIB
supergravity spectrum as the worldsheet theory reduces to a free
massive theory in the limit. In the second case, strings with
$p_{-}\to 0$ explore the UV limit of the worldsheet theory, which
gives the IR limit of the spacetime theory. In this limit we
expect to obtain the IIB supergravity spectrum but, as we will
see, we do not obtain all of it. In the limit $p_{-} \to 0$ the
supergravity part of the string spectrum comes from states with no
kinks or antikinks, and no oscillation modes from the free part of
the light-cone gauge string lagrangian. However, since there are
no fermionic zero modes corresponding to the interaction part of
the light-cone gauge string lagrangian, the Clifford algebra
satisfied by the GS fermionic zero modes is degenerate. This is
different from the usual non-degenerate Clifford algebra in the
case of flat, usual pp-wave string theories or our case when the
limit $\alpha'\to 0$ is considered. In the case where the
superpotential is for only one of the four complex transverse
coordinates of the string, the supergravity part of the string
spectrum consists of 128 instead of 256 states corresponding to
IIB supergravity. This background has an $SO(3)$ global symmetry.
Under this symmetry we find the space-time fields associated to
the supergravity part of the string spectrum. Some IIB
supergravity modes are absent when $p_{-} \to 0$, essentially
because the world-sheet system is at strong coupling, far from the
free-particle limit of \cite{metsaev1,metsaev2}.

There has been a fair amount of related work on pp-wave type
backgrounds. Such backgrounds with non-trivial R-R fluxes or other
fields, which can still be solved exactly at least in some cases,
have been analyzed recently in \cite{russo3,pope,russo2,blau2}. In
general these backgrounds preserve 16 supersymmetries and the
light-cone gauge string lagrangian contains harmonic oscillatory
type terms in the fields. The question of the amount of surviving
supersymmetry in D-brane configurations on pp-wave type
backgrounds found in \cite{malda1} has been addressed in
\cite{hikida}. A related background with localized D-branes
solutions on general pp-wave backgrounds has been found in
\cite{alday}. Other backgrounds producing interacting light-cone
gauge string models have been considered in the past
\cite{tseytlin2,klimcik} and more recently in
\cite{russo1,bakas,kim,bonelli}.

In section 2 we briefly review the general pp-wave solutions of
type IIB supergravity found in \cite{malda1}. In section 3 we find
the eigenvalues of the light-cone gauge string hamiltonian in the
background corresponding to the superpotential $Z^3-Z$. The
supergravity part of superstring spectrum as well as its
space-time interpretation is analyzed in section 4. In section 5
we discuss other cases when there are 2,3 or 4 interacting
theories. Also, we mention the background leading to ${\cal N}=2$
supersymmetric sine-Gordon theory on the worldsheet. Finally, we
present our conclusions in section 6.

\section{Superstrings on general IIB pp-wave backgrounds}

Maximally-supersymmetric pp-wave solutions of IIB supergravity
were considered in \cite{blau}. Strikingly, the world-sheet theory
is massive, but since the fields are free, the eigenvalues of the
light-cone superstring Hamiltonian can still be found
\cite{metsaev1,metsaev2}. The type IIB supergravity equations also
admit more general pp-wave type solutions of the form:
\begin{equation}
ds^{2}=-2dx^{+}dx^{-}+H(x^{i})dx^{+}dx^{+}+g_{ij}dx^{i}dx^{j}
\nonumber
\end{equation}
\begin{equation}
F_{5}=dx^{+}\wedge \varphi _{4}(x^{i}),
\end{equation}
where $x^{i}, i=1,...,8$ are the 8 transverse coordinates,
$x^{\pm} =\frac{1}{\sqrt{2}}(x^{0}\pm x^{9})$ and $\varphi _{4}$ a
four form in the transverse space. The transverse space can be
curved. All other fields are zero except the dilaton, which is
constant. The equations of motion of type-IIB supergravity imply
that the transverse space is Ricci flat, the five-form $F_{5}$ is
self-dual and that the four-form $\varphi_{4}$ and function $H$
are related by:
\begin{equation}
\nabla^{2}H=-32\left |\varphi \right |^{2}, \quad \quad \ast
_{10}F_{5}=F_{5}, \quad \quad R_{ij}[g]=0
\end{equation}
where $\left |\varphi \right |^{2}=\frac{1}{4!}\varphi
_{\mu\nu\lambda\delta}\varphi^{\mu\nu\lambda\delta}$, and
$\nabla^{2}$ is the laplacian in the transverse space. Since there
are no other forms $F_{5}$ is closed which along with the
self-duality of $F_{5}$ implies $\ast \varphi=-\varphi$ and
$d\varphi=0$. In complex coordinates the general 4-forms can be
classified by the $(a,b)$, the number of holomorphic and
antiholomorphic indices, with $a+b=4$. The anti-self dual 4-forms
are either $(1,3)$, their complex conjugates $(3,1)$, or $(2,2)$,
all of which can be written explicitly as
\begin{equation}
\varphi_{mn}=\frac{1}{6}\varphi_{m\bar{i}\bar{j}\bar{k}}
\epsilon^{\bar{i}\bar{j}\bar{k}\bar{n}}g_{n\bar{n}}, \quad \quad
\varphi_{l\bar{m}}=\frac{1}{2}g^{s\bar{s}}\varphi_{l\bar{m}s\bar{s}},
\end{equation}
where $\varphi_{mn}$ is a symmetric matrix and
$\varphi_{l\bar{m}}$ is a hermitian traceless matrix.

%These backgrounds
%preserve less supersymmetry, but their classification in terms of
%the amount of supersymmetry preserved has not been found.

The amount of supersymmetry preserved by this pp-wave type
background was studied in \cite{malda1}. The supersymmetry
transformations are generated by a chiral spinor $\epsilon$ with
16 complex components. Under the embedding $SO(1,1)\times
SO(8)\subset SO(9,1)$ the spinor $\epsilon$ is decomposed in
positive and negative $SO(1,1)$ and $SO(8)$ chiralities
($\Gamma's$ are $32\times 32$ matrices):
\begin{equation}
\epsilon=-\frac{1}{2}\Gamma_{+}\Gamma_{-}\epsilon-
\frac{1}{2}\Gamma_{-}\Gamma_{+}\epsilon
\equiv\epsilon_{+}+\epsilon_{-}
\end{equation}
The amount of supersymmetry preserved is obtained by solving the
Killing spinor equation. To do this it is convenient to work in
complex transverse space
$z^{i}=\frac{1}{\sqrt{2}}(x^{i}+ix^{i+4}),$ with $i=1,2,3,4.$
However, later in this paper when we analyze strings on these IIB
supergravity backgrounds we switch back to real coordinates.

Two classes of solutions have been found in \cite{malda1}, for
both flat and curved transverse space. The first class preserves
at least four spacetime supersymmetries and the second at least
two. For the flat case, which we consider in this paper, these
solutions can be parametrized by a holomorphic function $W$ and a
$4\times 4$ hermitian traceless constant matrix
$\varphi_{j\bar{k}}.$ The Killing spinors are:
\begin{equation}
\epsilon_{+}=\alpha |0\rangle+\zeta |\tilde{0}\rangle, \nonumber
\end{equation}
\begin{equation}
\epsilon_{-}=2i\Gamma_{-}\left(\zeta
\partial_{\bar{k}}\overline{W}-\alpha
\varphi_{j\bar{k}}z^{j}\right)\Gamma^{\bar{k}}|0\rangle+2i\Gamma_{-}\left(\alpha
\partial_{k}W-\zeta
\varphi_{k\bar{j}}\bar{z}^{\bar{j}}\right)\Gamma^{k}|\tilde{0}\rangle,
\end{equation}
where $\alpha$ and $\zeta$ are constant complex parameters. In the
Fock space formalism the vacuum $|0\rangle$ and the filled state
$|\tilde{0}\rangle=\frac{1}{4}\Gamma^{\bar{1}}\Gamma^{\bar{2}}
\Gamma^{\bar{3}}\Gamma^{\bar{4}}|0\rangle$ are given by:
\begin{equation}
\Gamma_{+}|0\rangle=\Gamma^{n}|0\rangle=0,\quad
\Gamma_{+}|\tilde{0}\rangle=\Gamma^{\bar{n}}|\tilde{0}\rangle=0,
\quad n=1,2,3,4
\end{equation}
In terms of $W$ the metric and the non-constant matrices
$\varphi_{mn}$ and $\varphi_{\bar{m}\bar{n}}$ are:
\begin{equation}
ds^{2}=-2dx^{+}dx^{-}-32(\left |\partial _{i}W \right
|^{2}+|\varphi_{j \bar{i}}z^{j}|^{2})dx^{+}dx^{+}+2dz^{i}d
\bar{z}^{\bar{i}} \nonumber
\end{equation}
\begin{equation}
\varphi_{mn}=\partial_{m}\partial_{n}W, \quad
\varphi_{\bar{m}\bar{n}}=\partial_{\bar{m}}\partial_{\bar{n}}\overline{W}
\label{back1}
\end{equation}
The second class of solutions, preserving at least two
supersymmetries, are parametrized by a real harmonic function $U$.
The Killing spinors are given by:
\begin{equation}
\epsilon_{+}=-\zeta |0\rangle+\zeta |\tilde{0}\rangle, \nonumber
\end{equation}
\begin{equation}
\epsilon_{-}=2i\Gamma_{-}\zeta \partial_{\bar{k}}U
\Gamma^{\bar{k}}|0\rangle-2i\Gamma_{-}\zeta\partial_{k}U
\Gamma^{k}|\tilde{0}\rangle
\end{equation}
The metric and the matrices $\varphi$ are:
\begin{equation}
ds^{2}=-2dx^{+}dx^{-}-32|\partial_{k}U|^{2}dx^{+}dx^{+}+2dz^{i}d\bar{z}^{\bar{i}},
\nonumber
\end{equation}
\begin{equation}
\varphi_{mn}=\partial_{m}\partial_{n}U, \quad
\varphi_{\bar{m}\bar{m}}=\partial_{\bar{m}}\partial_{\bar{n}}U,
\quad \varphi_{l\bar{m}}=\partial_{l}\partial_{\bar{m}}U
\end{equation}

In the rest of this paper, we consider only the first class of the
above two sets of solutions.  When strings are considered on the
above pp-wave type backgrounds, the light-cone gauge can be
implemented \cite{horowitz}.  We choose the light-cone gauge by
setting $X^{+}=\tau$, and
$\Gamma^{+}\theta=\Gamma^{+}\tilde{\theta}=0,$ where $\tau$ is the
worldsheet time and $\theta,\tilde{\theta}$ the two Majorana-Weyl
spinors. Since $(\Gamma^{+})^{2}=0$ they become the usual
Green-Schwarz $SO(8)$ spinors $S,\tilde{S}.$ In the usual
light-cone gauge procedure, $p_{-}$ is conserved and the
light-cone gauge hamiltonian is $H=-p_{+}$. The remaining killing
spinors $\epsilon_+$ are those not annihilated by $\Gamma^{+}$.
These spinors survive as linearly-realized supersymmetries on the
worldsheet. When all Ramond-Ramond fields are zero, the
superstring action becomes an ${\cal N}=(2,2)$ supersymmetric
non-linear sigma model written in terms of a K\"{a}hler potential.
When the R-R five-form field strength $F_{5}$ is turned on
(background (\ref{back1})), one expects to still have a
supersymmetric theory on the worldsheet, due to the preserved
spinors $\epsilon_{+}.$ As shown in \cite{malda1}, when the forms
$(1,3)$ and $(3,1)$ are present one can add an arbitrary
superpotential $W$ so that the worldsheet action in terms of the
superfields is:
\begin{equation}
S=\frac{1}{4\pi\alpha'}\int
d\tau\int^{2\pi\alpha'|p_{-}|}_{0}d\sigma(L_{K}+L_{W}) \nonumber
\end{equation}
\begin{equation}
L_{K}+L_{W}=\int d^{4}\theta
g_{i\bar{j}}\Phi^{i}\bar{\Phi}^{\bar{j}}+\frac{1}{2}\left(\int
d^{2}\theta W(\Phi^{i})+c.c.\right), \label{lagr}
\end{equation}
where $\Phi^{i}$ are chiral superfields with $i=1,2,3,4.$ An
interesting feature of this action is that it describes an
interacting field theory. In complex coordinates we have the
subgroup $SU(4)$ of $SO(8)$, and a negative chirality $SO(8)$
spinor decomposes in spinors with vector indexes as
$\textbf{8}_{-}\rightarrow \textbf{4}+\bar{\textbf{4}}$. The
transverse space is a Ricci-flat K\"{a}hler manifold. Therefore
there exists a covariantly constant spinor, and a general spinor
field can be expanded in terms of the ``one-particle states"
$\Gamma_{i}\eta_{0}$, ``two-particle states"
$\Gamma_{i}\Gamma_{j}\eta_{0}$, and so on, plus their complex
conjugates. In the case of negative chirality spinors
$S,\tilde{S}$ this decomposition defines the worldsheet spinors as
follows:
\begin{equation}
S=\psi^{i}_{-}\Gamma_{i}\eta_{0}+\bar{\psi}^{\bar{i}}_{-}\Gamma_{\bar{i}}\eta^{\ast}_{0}
\nonumber
\end{equation}
\begin{equation}
\tilde{S}=\psi^{i}_{+}\Gamma_{i}\eta_{0}+\bar{\psi}^{\bar{i}}_{+}
\Gamma_{\bar{i}}\eta^{\ast}_{0}, \label{spinors}
\end{equation}
where $\eta_{0}$ is a covariantly constant spinor annihilated by
all $\Gamma_{\bar{i}}.$

One can explicitly derive the action (\ref{lagr}) in the manner of
\cite{russo1}, where in the case of a non-constant R-R three-form
field $F_{3}$, a non-supersymmetric theory has been obtained on
the worldsheet. In our case the starting points are the background
(\ref{back1}) and the light-cone gauge Green-Schwarz lagrangian
when R-R fields are present, which can be found in
\cite{russo1,pope}. The part of the metric proportional to $\left
|\partial _{i}W \right |^{2}$ turns into an interacting term for
the bosonic fields on the worldsheet. The part of the lagrangian
connected to $F_{5}$ is proportional to
$iF_{+i_{1}i_{2}i_{3}i_{4}}\theta
\Gamma^{-}\Gamma^{i_{1}i_{2}i_{3}i_{4}} \tilde{\theta}.$ Choosing
a convenient representation of the gamma matrices to solve the
light-cone gauge conditions
$\Gamma^{+}\theta=\Gamma^{+}\tilde{\theta}=0,$ for $S,\tilde{S}$,
and then using the decomposition in (\ref{spinors}), the above
term of the light-cone lagrangian turns into Yukawa terms on the
worldsheet. Putting everything together, one then arrives at the
action (\ref{lagr}) when the forms $(1,3)$ and $(3,1)$ are turned
on. The arbitrary holomorphic function $W$, which parametrizes the
background, becomes the superpotential on the worldsheet.

A crucial question is whether the backgrounds reviewed in this
section describe solutions of superstring theory. Any action in
light-cone IIB GS formalism can be turned, by completing with a
proper set of fields, into a theory which is classically ${\cal
N}=(2,2)$ superconformally invariant.  If this new action is also
${\cal N}=(2,2)$ superconformally invariant at the quantum level,
the original light-cone GS background describes a solution of
superstring theory. It has been shown \cite{berk} by using the
above procedure, known as the $U(4)$ formalism, that the two
classes of IIB backgrounds found in \cite{malda1} are exact
solutions of superstring theory when the transverse target space
is flat.

\section{Superstring theory with superpotential $z^{3}-z$}

We have seen that superstrings in the above backgrounds produce,
in light-cone gauge, ${\cal N}=(2,2)$ supersymmetric interacting
field theories on the worldsheet. In general the action of an
${\cal N}=2$ supersymmetric theory contains in addition to the
part in (\ref{lagr}) an extra term $L_{V}$ which depends on a
holomorphic Killing vector $V.$ Including this term and setting
$F^{i}=-\frac{1}{2}g^{i\bar{i}}\partial_{\bar{i}}\bar{W}(\bar{Z})$
by using the equations of motion, a general ${\cal N}=(2,2)$
supersymmetric Lagrangian is given, in terms of the component
fields of the superfields by:
\begin{eqnarray}
L&=&L_{K}+L_{W}+L_{V} \nonumber \\
&=&g_{i\bar{j}}(\partial_{\tau}Z^{i}\partial_{\tau}\bar{Z}^{\bar{j}}-
\partial_{\sigma}Z^{i}\partial_{\sigma}\bar{Z}^{\bar{j}})+
ig_{i\bar{j}}\bar{\psi}^{\bar{j}}_{+}(\partial_{\tau}-\partial_{\sigma})\psi^{i}_{+}+
ig_{i\bar{j}}\bar{\psi}^{\bar{j}}_{-}(\partial_{\tau}+\partial_{\sigma})\psi^{i}_{-} \nonumber \\
&+&\frac{i}{2}\partial_{i}\partial_{j}W(Z)\psi^{i}_{+}\psi^{j}_{-}+
\frac{i}{2}\partial_{\bar{i}}\partial_{\bar{j}}\bar{W}
(\bar{Z})\bar{\psi}^{\bar{i}}_{-}\bar{\psi}^{\bar{j}}_{+}-
\frac{1}{4}g^{i\bar{j}}\partial_{i}W(Z)\partial_{\bar{j}}\bar{W}(\bar{Z}) \nonumber \\
&-&\left|\tilde{m}\right|^{2}g_{i\bar{j}}V^{i}V^{\bar{j}}-
\frac{i}{2}(g_{i\bar{i}}\partial_{j}V^{i}-
g_{j\bar{j}}\partial_{\bar{i}}V^{\bar{j}})(\tilde{m}\bar{\psi}^{\bar{i}}_{-}\psi^{j}_{+}
+\bar{\tilde{m}}\bar{\psi}^{\bar{i}}_{+}\psi^{j}_{-}),
\label{generalaction}
\end{eqnarray}
where the last line represents the extra term $L_{V}$, and the
superfields are expanded as usual: $
 \Phi^{i}=Z^{i}+\sqrt{2}\theta^{+}\psi^{i}_{+}+\sqrt{2}\theta^{-}\psi^{i}_{-}+
 2\theta^{+}\theta^{-}F^{i}+... .$
Since we have closed strings the worldsheet is a cylinder. The
periodicity of the Green-Schwarz fermions leads to periodic
boundary conditions on the worldsheet fermions.

In general these two-dimensional theories are massive and, in our
case, they are defined on a cylinder with periodic boundary
conditions on all fields. If these theories on the worldsheet are
integrable, it should be possible at least in principle to find
the string spectrum. A simplification is to take a superpotential
of the form
$W=W_{1}(Z^{1})+W_{2}(Z^{2})+W_{3}(Z^{3})+W_{4}(Z^{4})$ so that we
obtain four independent two-dimensional theories. We consider the
massive integrable theory given by the superpotential $W=\lambda
((Z^{1})^{3}/3- \delta^{2}Z^{1})$, where $Z^{1}$ is the
superstring coordinate in the 1-complex direction and
$\delta,\lambda$ real parameters. To find the eigenvalues of the
string light-cone gauge hamiltonian we need to analyze the full
worldsheet theory, while to find the supergravity part of the
string spectrum we have to consider either the $\alpha'\to 0$ or
$p_{-} \to 0$ limit of the worldsheet theory, which we will do in
section 4. In 2d language, this integrable ${\cal N}=(2,2)$
supersymmetric field theory is a relevant perturbation of the
first superconformal minimal model \cite{fendley3}. One can
bosonize the fermions to obtain the ordinary sine-Gordon model at
a special coupling where it is supersymmetric, but we will
continue to use the manifestly-supersymmetric language.

In superstring theory we take the transverse space to be flat:
$g_{i\bar{j}}=\delta_{i\bar{j}}.$ As shown in \cite{malda1}, in
flat transverse space we have a holomorphic vector field $V$, and
it is related to $\varphi_{i\bar{j}}$ as:
\begin{equation}
V_{i}=-i\varphi_{i\bar{j}}\bar{Z}^{\bar{j}},\quad
V_{\bar{j}}=i\varphi_{i\bar{j}}Z^{i}
\end{equation}
However, we take the $(2,2)$ forms zero, $\varphi_{i\bar{j}}=0$,
so that $V=0$. For the above superpotential the metric and the
anti-self-dual 4-form are:
\begin{equation}
ds^{2}=-2dx^{+}dx^{-}-32\lambda^{2}(z^{1}z^{1}-
\delta^{2})(\bar{z}^{\bar{1}}\bar{z}^{\bar{1}}-
\delta^{2})dx^{+}dx^{+}+2dz^{i}d \bar{z}^{\bar{i}} \nonumber
\end{equation}
\begin{equation}
F_{5}=dx^{+}\wedge \varphi_{4} \quad \varphi_{4}=(2\lambda
z^{1})dz^{1}\wedge d\bar{z}^{\bar{2}}\wedge
d\bar{z}^{\bar{3}}\wedge d\bar{z}^{\bar{4}}+(2\lambda
\bar{z}^{\bar{1}})d\bar{z}^{\bar{1}}\wedge dz^{2}\wedge
dz^{3}\wedge{dz^{4}},  \label{background}
\end{equation}
Since we take $W$ to depend only on $Z^{1}$, the other theories in
$Z^{2},Z^{3},Z^{4}$ are massless and free and they can be easily
solved. For this part one has the usual light-cone gauge
hamiltonian consisting of a kinetic part and an oscillatory part
expressed in terms of rising and lowering bosonic and fermionic
operators:
\begin{equation}
H_{1}=\frac{1}{2|p_{-}|}\sum_{\alpha}^{}p^{\alpha}p^{\alpha}+
\frac{1}{\alpha'|p_{-}|}(N+\tilde{N}),
\end{equation}
where the summation is over the free coordinates
$\alpha=2,3,4,6,7,8.$ The occupation numbers expressed in complex
coordinates are:
\begin{equation}
N=\sum_{n=1}^{\infty}\left(\alpha^{\bar{k}}_{-n}\alpha^{k}_{n}+
\alpha^{k}_{-n}\alpha^{\bar{k}}_{n}+nd^{\bar{k}}_{-n}d^{k}_{n}+
nd^{k}_{-n}d^{\bar{k}}_{n}\right), \nonumber
\end{equation}
\begin{equation}
\tilde{N}=\sum_{n=1}^{\infty}\left(\tilde{\alpha}^{\bar{k}}_{-n}\tilde{\alpha}^{k}_{n}+
\tilde{\alpha}^{k}_{-n}\tilde{\alpha}^{\bar{k}}_{n}+
n\tilde{d}^{\bar{k}}_{-n}\tilde{d}^{k}_{n}+
n\tilde{d}^{k}_{-n}\tilde{d}^{\bar{k}}_{n}\right)
\end{equation}
As usual for a free hamiltonian the ground state of $H_{1}$ is
defined by:
\begin{equation}
\alpha^{k}_{n}|0\rangle=\alpha^{\bar{k}}_{n}|0\rangle=\tilde{\alpha}^{k}_{n}|0\rangle=
\tilde{\alpha}^{\bar{k}}_{n}|0\rangle=
d^{k}_{n}|0\rangle=d^{\bar{k}}_{n}|0\rangle=\tilde{d}^{k}_{n}|0\rangle=
\tilde{d}^{\bar{k}}_{n}|0\rangle=0 \quad n>0, \quad k=2,3,4
\nonumber
\end{equation}

To find the total light-cone string hamiltonian $H=H_{1}+H_{2}$ we
need to find the spectrum of the interacting part $H_{2}$, with
Lagrangian given by (\ref{generalaction}) with $W=\lambda
\left(\frac{1}{3}(Z^{1})^{3}- \delta^{2}Z^{1}\right)$ When
$\lambda=0$ the theory becomes the flat space type IIB
superstring. We are interested in the superstring theory in the
general pp-wave background with $\lambda\neq 0$ when at least four
space-time supersymmetries are preserved. The parameter $\lambda$
can be absorbed into rescaling of $x^{+},x^{-}$ which corresponds
to rescaling of light-cone gauge energy and $|p_{-}|$; so in fact
$\lambda$ can be set to a fixed nonzero value. Therefore the
eigenvalue configuration of $H$ depends on $\delta$, but not
$\lambda$ (other than just rescaling of energy). This is different
from the maximally supersymmetric pp-wave background where any
particular value of the background parameter is not special due to
rescaling of $x^{+},x^{-}.$ It is convenient to rewrite the
corresponding two-dimensional ${\cal N}=(2,2)$ supersymmetric
field theory in terms of canonical dimensionless fields. After
re-scaling $Z\rightarrow \sqrt{2\pi\alpha'}Z, \psi\rightarrow
\sqrt{2\pi\alpha'}\psi,\delta\rightarrow\sqrt{2\pi\alpha'}\gamma,
\lambda\rightarrow \frac{\mu}{\sqrt{2\pi\alpha'}}$, we have
$\gamma$ dimensionless and $\mu$ a mass parameter. The action is
then
$$
S_{2}=\frac{1}{2}\int d\tau \int_{0}^{2\pi\alpha'|p_{-}|}d \sigma
L_{2}
$$
\begin{eqnarray}
L_{2}&=&\partial_{\tau}Z
\partial_{\tau}\bar{Z}-
\partial_{\sigma}Z \partial_{\sigma}\bar{Z}-
\frac{1}{4}\mu^{2}(Z^{2}-\gamma^{2})(\bar{Z}^{2}-\gamma^{2}) \nonumber \\
&+&i\bar{\psi}_{+}(\partial_{\tau}-\partial_{\sigma})\psi_{+}+
i\bar{\psi}_{-}(\partial_{\tau}+\partial_{\sigma})\psi_{-}+i \mu
Z\psi_{+}\psi_{-}+i\mu \bar{Z}\bar{\psi}_{-}\bar{\psi}_{+},
\label{interact}
\end{eqnarray}

On the infinite line, the low-lying energies of many 2d
Lorentz-invariant field theories are written as $E=\sqrt{p^2 +
m^2}$, where $m$ is the mass of a particle in the theory, and $p$
its momentum. If the particles are interacting, the energy of
multi-particle states cannot be generically determined, but the
qualitative behavior can be described in terms of particles.
Because of the non-renormalization theorems in an ${\cal N}=(2,2)$
theory, the particle content of the theory (\ref{interact}), can
be inferred directly from the superpotential
\cite{fendley3,fendley2}.  The infinite-volume spectrum is
described in terms of kinks interpolating between the vacua; each
kink is a supersymmetry doublet of fermion numbers $1/2$ and
$-1/2$. In this model there are no breathers; but they exist in
the super sine-Gordon case, which we will discuss later. BPS kinks
have mass $m=|W(Z(\sigma\to\infty))- W(Z(\sigma\to -\infty))|$
where $Z(\sigma\to \pm\infty)$ are the values of the bosonic field
at spatial infinity for a kink in infinite volume. For the
superpotential $W=\mu \left(\frac{Z^{3}}{3}- \gamma^{2}Z\right)$
in (\ref{interact}) the two vacua are at $Z=\pm\gamma$, so $m =
4\mu \gamma^{3}/3 =2\lambda \delta^3/(3\pi\alpha')$.

On a compact space, such as the one we have with circumference
$R=2\pi\alpha' |p_{-}|$, finding the spectrum is a more difficult
problem. Even if the kinds of particles in the spectrum are known,
their mass depends on $R$; in Feynman diagram language this is
because virtual particles can propagate ``around the world''
\cite{Luscher}. The spectrum thus depends on the boundary
conditions. Here we take both the bosonic and fermionic fields to
satisfy periodic boundary conditions
$Z(\tau,\sigma)=Z(\tau,\sigma+R),
\psi(\tau,\sigma)=\psi(\tau,\sigma+R)$. As mentioned in
\cite{bakas} and \cite{malda1}, besides considering strictly
periodic boundary conditions one may also impose identifications
in the space of field configurations. This will modify the
spectrum of the theory. At this point we can consider two
situations. One is given by taking the identification $Z\simeq -Z,
\psi\simeq -\psi$ in the space of fields. This means that the
transverse space in the string light-cone gauge formalism is
$\mathbb{C}/\mathbb{Z}_{2}\times \mathbb{C}^{3}.$ The other
situation to consider, and the one we deal with in most of this
paper, is the one with no identifications in field configuration,
and the transverse space in string theory is then
$\mathbb{C}^{4}\simeq \mathbb{R}^{8}$. The requirement of
translation invariance along $\sigma$ gives the restriction on
physical states:
\begin{equation}
P_{2}|\psi_{phys}\rangle=\frac{N-\tilde{N}}{\alpha'
|p_{-}|}|\psi_{phys}\rangle
\end{equation}

Finding the eigenvalues of the $H_2$ is equivalent to finding the
energy levels of the $1+1$-dimensional field theory with
Lagrangian (\ref{interact}) on a circle of circumference $R$. In
the world-sheet theory, the only two dimensionful quantities are
the mass $m$ of the kinks, and the size of the system $R$.  On
dimensional grounds, the energy can therefore be written in terms
of a scaling function: $E = f(mR)/R$. This means in terms of
string quantities, the excited-state energies of the interacting
part $H_{2}$ of the light-cone Hamiltonian are
\begin{equation}
E_{2}^{(i)}=\frac{1}{12\alpha'
|p_{-}|}f^{(i)}\left(\frac{4}{3}\lambda \delta^{3}|p_{-}|\right)
\label{exenergies}
\end{equation}
where the scaling function $f^{(i)}$, with $i=1\dots\infty$, can
be computed using the 1+1-dimensional integrability techniques.
The above expression is valid for any $\mu \neq 0$ or in string
scale $\lambda \neq 0$. Note that regardless of the coupling no
breather states exist, i.e. no small fluctuation modes at the
minima of the superpotential exist, as one would expect. However,
in the point particle limit $\alpha' \to 0$ with $\lambda, \delta$
and $p_{-}$ fixed, the worldsheet theory reduces to a free massive
theory with mass $\lambda \delta$ (as $\mu \to 0, \gamma \to
\infty$ with $\mu \gamma =\lambda\delta=fixed$). This mismatching
between the existence and non-existence of the fluctuation modes
at the minima of the superpotential in the $\mu \to 0$ case and
$\mu \neq 0$ case respectively, signals that the two situations
are infinitely far (in the renormalization-group sense). We will
discuss more this issue in Section 4.

On general grounds, we know the ground state of $H_{2}$ with
periodic boundary conditions has zero energy $E_{2}^{(0)}=0$. In
this case the Witten index \cite{witten} is two; there are two
bosonic ground states with $P_{2}=0$ and no fermionic zero modes.
This is fairly obvious from the superpotential, whose bosonic part
has two minima. Since $N=\tilde{N}=0$ the ground state of the
light-cone gauge hamiltonian $H=H_{1}+H_{2}$ contains no
oscillators from the free part $H_{1}$. Therefore the ground state
energy of $H$ is just the kinetic part
$\frac{1}{2|p_{-}|}\sum_{\alpha}^{}p^{\alpha}p^{\alpha}.$ As usual
there is a degeneracy associated with the fermionic zero modes
from the free part $H_{1}$, but now there is also a bosonic
degeneracy coming from interacting part $H_{2}$. However, in
contrast with the flat space now there are fewer fermionic zero
modes as there are no fermionic zero modes for the ground state of
$H_{2}$. There are two ways to take the string's point particle
limit: $p_{-} \to 0$ or $\alpha' \to 0$. It appears that here the
two limits give different answers in contrast to flat space or the
maximally supersymmetric pp-wave. In the first case when $p_{-}
\to 0$ the expression (\ref{exenergies}) is valid in the limit
$p_{-} \to 0$, all excited levels decouple and we are left with
the two bosonic ground states discussed above. Physically these
states correspond to a point particle fixed at either
$x^{1}=\delta, x^{5}=0$ or $x^{1}=-\delta, x^{5}=0$, but free in
the other transverse directions
$x^{2},x^{3},x^{4},x^{6},x^{7},x^{8}.$ In the other point particle
limit, $\alpha' \to 0$, we have a free massive theory on the
worldsheet and we are left with an infinite number of states which
do not decouple in the limit. Physically the string reduces to a
point particle in a harmonic potential $(X^{1})^{2}+(X^{5})^{2}$
at either $x^{1}=\delta, x^{5}=0$ or $x^{1}=-\delta, x^{5}=0$, but
free in the other transverse directions. In the next section we
will discuss the point particle string spectrum in more detail.

Finding the rest of the finite-size spectrum is a much more
difficult problem.  However, some progress has been made in
integrable field theories by using the thermodynamic Bethe ansatz.
Usually this technique is used to compute the free energy for a
system on the infinite line at some non-zero temperature $T$. This
amounts to computing the partition function on an infinitely-long
cylinder of circumference $R=1/T$, which yields the Casimir energy
of the system in size $R$. With periodic boundary conditions
around the cylinder, the Casimir energy here is exactly $0$, as we
have already indicated.  In some cases, however, one can continue
the thermodynamic Bethe ansatz equations to obtain the exact
finite-size energies of excited states \cite{BLZ}. One ends up
with a set of coupled non-linear integral equations whose solution
yields the energies. For the above action $S_{2}$ with periodic
boundary conditions, this computation has already been done for
states with total momentum $P_{2}=0$ \cite{fendley4}. This
analysis has been extended to find the $R$-dependence of
single-particle states as well \cite{balog}; these one-particle
states can appear when one takes the identifications in transverse
target space $Z\simeq -Z, \psi\simeq -\psi.$ In what follows we
consider no identifications in the transverse space.

Since the analysis of \cite{fendley4} is quite technical, we do
not reproduce it here, but refer the reader there for details.  In
the $mR \to 0$ limit, the energy levels of the states found in
\cite{fendley4} are
$E_{2}^{(j)}=\frac{\pi}{6R}\left[(2j+1)^2-1\right]$,
$j=1\dots\infty$. In the limit when $mR$ is very large, the
particles are far apart and effectively free.  Because of the
periodic boundary conditions, the energy levels correspond to
states with the same number of kinks and antikinks, each of mass
$m$.  For example, the first excited level in the large $mR$ limit
consists of two free particles with momenta $\pm \pi/R$ and total
energy $E_{2}^{(1)}\approx 2\sqrt{m^{2}+(\pi/R)^{2}}$. The second
excited state $E_{2}^{(2)}$ also represents two free particles,
but now with momenta $\pm 2\pi/R.$ In the large-$mR$ limit, it was
shown in \cite{fendley4} that the energies $E_{2}^{(3k+1)}$ and
$E_{2}^{(3k+2)}$, $k=0\dots\infty$ approach each other; they
correspond to states with $2k+1$ kinks and $2k+1$ antikinks. The
energies $E_{2}^{(3k)}$, $k=1\dots\infty$ correspond to states
with $2k$ kinks and $2k$ antikinks.  These do not comprise all
excited-state energy levels, but it seems likely that the
techniques of \cite{fendley4,balog} could be extended to others if
desired.  Away from large $mR$ limit, one must take into account
the interactions between the kink and antikink. We have
numerically solved the non-linear integral equations found in
\cite{fendley4} for $E_{2}^{(1)},E_{2}^{(2)}, E_{2}^{(3)}$, and
displayed the results in figure 1.

We consider only excited states of the two-dimensional interacting
theory which have zero momentum $P_{2}=0$, so that the physical
states of light-cone string hamiltonian $H$ satisfy $N=\tilde{N}$.
Apart from the continuous part
$\frac{1}{2|p_{-}|}\sum_{\alpha}^{}p^{\alpha}p^{\alpha}$ the
light-cone gauge hamiltonian has a discrete spectrum. On the same
graph (figure 1) viewed for a constant value of $R$ we can also
represent the oscillatory part of $H_{1}$ depending only on $R$,
namely $\frac{2N}{\alpha' |p_{-}|}$ for $N=1,2,...$.
\begin{figure}[ht]
\centerline{\includegraphics[scale=0.6]{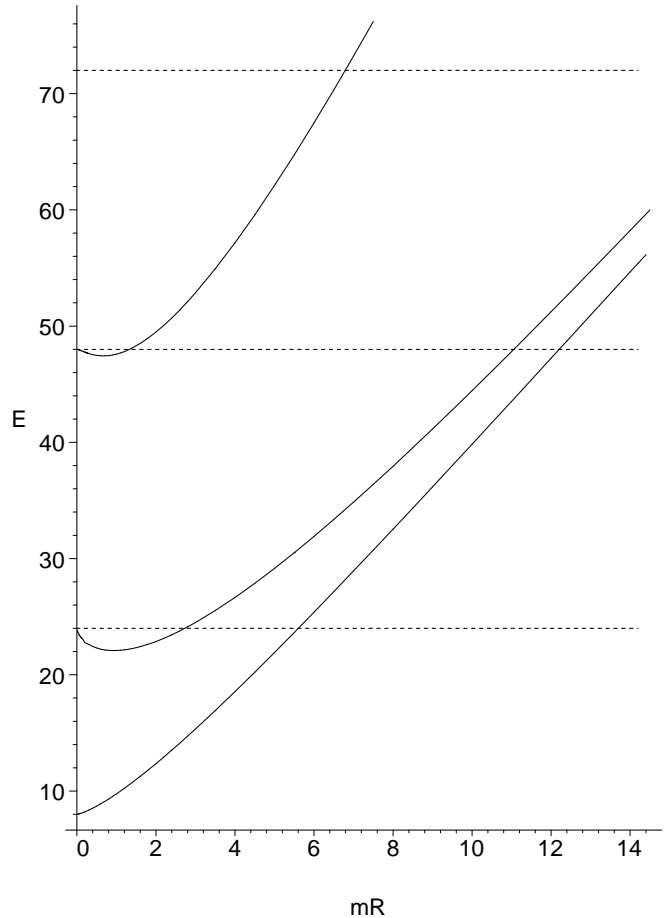}} \caption{The
first three excited-state energies of $H_{2}$:
$E_{2}^{(1)},E_{2}^{(2)},E_{2}^{(3)}$ represented by continuous
lines. The first three oscillator levels of $H_{1}$ represented by
dashed lines. The energies are in units of
$\frac{1}{12\alpha'|p_{-}|},$ while $mR$ in string scale is
$\frac{4}{3}\lambda \delta^{3}|p_{-}|.$}
\end{figure}
Thus all eigenvalues of the light-cone string hamiltonian $H$ can
be obtained. They depend on the background inputs $\lambda,\delta$
and on the string parameter $|p_{-}|.$ The eigenvalue
configuration depends non-trivially on $mR=\frac{4}{3}\lambda
\delta^{3}|p_{-}|.$ The intersections of $\frac{2N}{\alpha'
|p_{-}|}$ with the eigenvalues of $H_{2}$ gives sectors in $mR$;
in two different such sectors the eigenvalue configuration is
different. For example, apart from the non-oscillatory part of
$H_{1}$, the first six levels of $H$ for $2.7<mR<5.6$ are:
\begin{equation}
E^{(0)}=0, \quad E^{(1)}=E_{2}^{(1)}, \quad
E^{(2)}=\frac{2}{\alpha' |p_{-}|}, \quad E^{(3)}=E_{2}^{(2)},
\quad E^{(4)}=E_{2}^{(1)}+\frac{2}{\alpha' |p_{-}|}, \quad
E^{(5)}=\frac{4}{\alpha' |p_{-}|}
\end{equation}
while for $5.6<mR<11.1$ they are:
\begin{equation}
E^{(0)}=0, \quad E^{(1)}=\frac{2}{\alpha' |p_{-}|},\quad
E^{(2)}=E_{2}^{(1)}, \quad E^{(3)}=E_{2}^{(2)}, \quad
E^{(4)}=\frac{4}{\alpha' |p_{-}|}, \quad
E^{(5)}=E_{2}^{(1)}+\frac{2}{\alpha' |p_{-}|}
\end{equation}

Due to the fact that $g_{++}\neq 0$ the string feels a
gravitational force that pulls it to the regions where $-g_{++}$
is minimum, namely the regions with $z^{1}=\pm\delta$
($x^{1}=\pm\delta,x^{5}=0$). Then kink-antikink pairs on the
worldsheet going between minima $x^{1}=\pm\gamma, x^{5}=0$
correspond to a string that goes between the two positions where
$-g_{++}$ has a minimum in target space, namely $\pm\delta.$

As the (infinite-volume) kink mass $m$ gets bigger due to changes
in background parameters $\lambda,\delta$, the oscillatory states
of $H_{1}$ have lower energy than the massive states of $H_2$. As
$m\rightarrow\infty$ with $p_{-}$ fixed, the states of $H_2$
effectively decouple. In the string scale, this limit corresponds
to $\delta\rightarrow\infty$ and/or to a very strong R-R field
$\lambda\rightarrow\infty.$ Therefore in this limit the transverse
space seen by the string in its excited states is six-dimensional
and it consists in two hyperplanes sitting at $x^{1}=\delta,
x^{5}=0$ and $x^{1}=-\delta, x^{5}=0.$ In this limit the string is
highly excited in the free directions, and it is in the ground
state in the interacting directions. Alternatively, we can keep
the background fixed, i.e. $\lambda,\delta$ fixed. Then the above
limit is obtained for strings with very large $|p_{-}|$, which is
the IR limit of the worldsheet theory. It seems possible that in
this limit it might be a correspondence between the string theory
under discussion and a string theory on a partially discrete
target space, e.g.\ a transverse target space
$\mathbb{R}^{6}\times \left\{-\delta,\delta \right\}\times
\left\{0\right\}.$ Because of the UV/IR relationship between
worldsheet and target space scales, the above discrete space
appears at very short space-time scales.

\section{Supergravity part of the superstring spectrum}

We now return to the low-energy states of $H$ and analyze the
supergravity part of superstring, the ``massless" string modes.
The supergravity part of string spectrum comes from all possible
states giving light-cone energies which do not decouple in the
limit $\alpha'\rightarrow 0$ while keeping everything else fixed.
As we already mentioned, in the $\alpha' \to 0$ limit the
worldsheet theory becomes a free massive theory. In this case, as
in the case of the maximally supersymmetic pp-wave, we expect to
obtain the usual type IIB supergravity spectrum. As we have seen,
the string's point particle limits $\alpha' \to 0$ and $p_{-} \to
0$ give different answers. In the first case we obtain the IIB
supergravity spectrum, but in the second case some of these modes
are missing, as we will show in this section.

In the remaining of this section we analyze the supergravity
spectrum obtained in the limit $p_{-} \to 0$, which means the
worldsheet radius $R \to 0$, while everything else is kept fixed.
As opposed to the case of a free massive superfield on the world
sheet discussed in \cite{metsaev1,metsaev2}, only a few modes for
the the theory discussed in section 3 remain in the $p_{-} \to 0$
limit. To explain this issue a little further, we amplify on a
discussion in \cite{malda1}. A free massive superfield of mass $M$
on the world sheet of circumference $R$ results from the
superpotential $M z^2/2$. The energy levels are are $\sqrt{(k_n)^2
+M^2},$ where $k_n = 2n\pi/R$ is the world-sheet momentum. Thus in
the $R\to 0$ limit, all of these modes have very large energy
except for the zero-momentum modes with $n=0$. For a free massive
superfield, there are an infinite number of these, because one can
have as many zero-momentum bosonic modes present as desired.

The question to ask is then: what happens to these zero-momentum
modes when interactions are included? It is useful to answer this
question first in the case of super sine-Gordon, where the
superpotential is $W=(M/\beta^2)\cos(\beta z)$, with $M$ a mass
parameter and $\beta$ the usual coupling. In the limit $\beta\to
0$, this reduces to the free superfield, and we have an infinite
number of zero modes. This fits in perfectly with the results from
integrable field theory: in the limit $\beta\to 0$ there are an
infinite number of ``breather'' states in the spectrum, which have
masses $M, 2M, 3M,\dots$. The kinks have mass proportional to
$1/\beta$ in this limit, and so decouple.  Thus a state with a
single breather of momentum zero will have energy $jM$ for some
integer $j$. All of these states precisely reproduce the spectrum
of a free superfield (the analogous statement for the ordinary
sine-Gordon model is that $\beta\to 0$ spectrum is that of a
single free boson).

Since the super-sine-Gordon model is integrable for any $\beta$,
we can understand what happens to these states as one increases
$\beta$. What is known is that as $\beta$ is increased, the
breather states drop out of the spectrum. This does not mean that
there are discontinuities in the energy levels as a function of
$\beta$ -- it means that a (single-particle) breather state turns
into a (two-particle) kink-antikink state as $\beta$ is increased.
One can see this explicitly in the Bethe-ansatz solution of the
ordinary sine-Gordon/massive Thirring model \cite{hank}. For large
enough $\beta$ ($\beta^2=8\pi$ in the usual normalization of
super-sine-Gordon), {\em all} of the breather states have dropped
out of the theory. For ordinary sine-Gordon, this point is where
the sine-Gordon model is equivalent to a free Dirac fermion, so we
can easily see what has happened to the zero-momentum
single-particle breather states: they correspond to a fermion and
antifermion (kink and antikink in the sine-Gordon language) with
opposite momentum.  The crucial point is that because here the
excitations are fermions, there is only one state where both
particles have zero momentum: the state with one fermion of zero
momentum and one antifermion of zero momentum. This state can be
identified with the breather of lowest mass. The higher
zero-momentum breather states must therefore be identified with a
state with fermion and antifermion, but where the two particles
have non-zero individual momentum. In a finite-size system, these
states must have energy proportional to $1/R$, and so disappear in
the small-$R$ limit. It would be quite interesting to confirm this
by extending the computation of \cite{balog} to compute the energy
of the single-particle breather states in finite volume.

In the super-sine-Gordon case, the same thing must happen -- only
two states of relative and total momenta zero remain at
$\beta^2=8\pi$. (It is two here because there is a supersymmetric
doublet of kinks of charge $\pm 1/2$ \cite{pk2}; there are
therefore two two-particle states of charge $0$ and individual
momenta zero.) This follows from the fact that in all theories
solvable by the Bethe ansatz, the excitations are fermionic in
that only one excitation of a given type is allowed per level (for
example, in ordinary sine-Gordon, one can only have a single
zero-momentum breather of a given type present). For
super-sine-Gordon, this fact was confirmed by the thermodynamic
Bethe ansatz computation of \cite{pk2}. Therefore, all but two of
the zero-charge zero-momentum states present as $\beta \to 0$ must
have energy proportional to $1/R$ (possibly multiplied by a
logarithm) by the time $\beta^2$ reaches $8\pi$.

We have therefore argued that although there are an infinite
number of low-energy states in the limit of free particles on
world sheet, i.e. $\beta \to 0$ limit, only a small number of
states survive for strong enough interactions in the case of super
sine-Gordon. Let us see what happens when considering the
superpotential $z^3-z$ studied above. In the small-$R$ limit, i.e.
the UV limit on the worldsheet, the spectrum is that of the
conformal field theory with superpotential $\mu z^3$, or in string
scale $\lambda z^3$; with periodic boundary conditions this theory
has only two ground states. All other states will have energy
proportional to $1/R$. Another way of seeing how this can happen
is by deforming the free massive theory: if one adds a
superpotential $z^3$ to the $z^2$, one obtains the $z^3-z$ theory
(the two are equivalent by a constant shift in $z$). However, the
conformal field theory with $\mu z^3$ is infinitely far (in the
renormalization-group sense) from the free massive superfield;
this must be so because the theories have different Witten indices
\cite{pk3}. This means that the conformal field theory with
superpotential $\mu z^3$ cannot be reached by starting with a free
massive superfield and adding small perturbations.\foot{In
\cite{pk3} it is argued that a Landau-Ginzburg theory given by a
large class of superpotentials flows in the UV limit to a free
massless superfield. However, the free theory is reached
asymptotically, which means that in the UV region the theory is
not equivalent with a free CFT plus a small perturbation. The
superpotential does not go away completely even in the far UV,
which is a sign that non-perturbative effects remain in the
limit.} Thus it is not surprising that all but a finite number of
low-energy states go away in the small $R$-limit.

In fact, generically any worldsheet theory can effectively be
described in the $R\to 0$ limit by a conformal field theory.
Because the only mass scale here is $1/R$, all states other than
exact ground states will have high energy.  If the conformal
theory consists of free superfields, then the corresponding
supergravity spectrum will be the usual one. If the model is
solvable by the Bethe ansatz, then states will drop out of the
spacetime supergravity spectrum unless the worldsheet spectrum
contains an infinite number of particles. Moreover, it seems
likely generically, that states will drop out unless the
interactions in the worldsheet conformal field theory are weak. To
give a well-known example, in Calabi-Yau compactification, the
world-sheet theory is a sigma model with a Calabi-Yau manifold as
a target space; this can change the low-energy spectrum
dramatically.

We can see the effects of these considerations on our full theory
by examining the fermion zero-mode spectrum explicitly. The ground
state has a degeneracy coming from the two bosonic states of the
interacting 2d theory. In addition there is the degeneracy from
the algebra satisfied by the zero modes of the light-cone GS
fermions $S$ and $\tilde{S}$ both having the same chirality, which
we take to be negative. We need to find an irreducible
representation of this algebra. As we already mentioned there are
no fermionic zero modes in the interacting directions, and the
other worldsheet fermionic zero modes satisfy:
\begin{equation}
\left\{d_{0}^{i},d_{0}^{j}
\right\}=\left\{d_{0}^{\bar{i}},d_{0}^{\bar{j}}\right\}=\left\{\tilde{d}_{0}^{i},
\tilde{d}_{0}^{j}\right\}=\left\{\tilde{d}_{0}^{\bar{i}},\tilde{d}_{0}^{\bar{j}}\right\}=0
\nonumber
\end{equation}
\begin{equation}
\left\{d_{0}^{i},d_{0}^{\bar{j}}\right\}=\left\{\tilde{d}_{0}^{i},\tilde{d}_{0}^{\bar{j}}\right\}=\delta^{i\bar{j}},\quad
i,j=2,3,4 \label{zeromodes}
\end{equation}
The gamma matrices in complex coordinates satisfy
$[{\Gamma_{i},\Gamma_{\bar{i}}}]=2\delta_{i\bar{i}}.$ As usual in
complex space we also define
$\Gamma_{i\bar{i}}=\frac{1}{2}(\Gamma_{i}\Gamma_{\bar{i}}-
\Gamma_{\bar{i}}\Gamma_{i})$, with $i=1,2,3,4.$ They satisfy:
\begin{equation}
[\Gamma_{i\bar{i}},\Gamma^{\bar{k}}]=2\Gamma_{i}\delta^{\bar{k}}_{\bar{i}},
\quad
[\Gamma_{i\bar{i}},\Gamma^{k}]=-2\Gamma_{\bar{i}}\delta^{k}_{i}
\nonumber
\end{equation}
A constant spinor annihilated by all $\Gamma_{\bar{i}}$ can be
taken to be $\eta_{0}=\frac{i}{2}(-1,-1,-1,-1).$ Then
\begin{equation}
\Gamma_{1}\eta_{0}=\frac{i}{2}(1,-1,-1,-1), \quad
\Gamma_{\bar{1}}\eta_{0}^{\ast}=\frac{i}{2}(-1,1,1,1)
\end{equation}
and similar expressions for the other actions of $\Gamma_{i}$ and
$\Gamma_{\bar{i}}$ on $\eta_{0},\eta_{0}^{\ast}$. Using
(\ref{spinors}) one obtains the GS fermions $S$ and $\tilde S$ in
terms of the worldsheet fermions.
The light-cone gauge superstring lagrangian $L$, taken in the
string scale, is then
\begin{eqnarray}
L&=&\frac{1}{2}\left(\sum_{i=1}^{8}\partial_{+}X^{i}\partial_{-}X^{i}\right)-
\frac{\lambda^{2}}{4}\left[\frac{1}{4}\left((X^{1})^{2}+(X^{5})^{2}\right)^{2}+
\delta^{4}-\delta^{2}\left((X^{1})^{2}-(X^{5})^{2}\right)\right]
\nonumber \\
&+&\frac{i}{4}\left(S^{a}\partial_{+}S^{a}+\tilde{S}^{a}\partial_{-}\tilde{S}^{a}\right)
-\frac{i\lambda}{8\sqrt{2}}\left(X^{1}S^{a}M_{ab}\tilde{S}^{b}-X^{5}S^{a}P_{ab}\tilde{S}^{b}\right),
\label{lagran3}
\end{eqnarray}
where $\partial_{\pm}=\partial_{\tau}\pm \partial_{\sigma}$, and
matrices $M$ and $P$ are:
\begin{equation}
M=
\begin{pmatrix}%%{ll}
Q & 0 \\
0 & -Q
\end{pmatrix}
 \quad P=
\begin{pmatrix}%%{ll}
0 & Q \\
Q & 0
\end{pmatrix}
, \quad \hbox{with} \quad Q=
\begin{pmatrix}%%{llll}
\ 1&-1&-1&-1 \\
-1&1&1&1 \\
-1&1&1&1 \\
-1&1&1&1
\end{pmatrix}
,
\end{equation}
The background (\ref{background}) in real coordinates becomes:
\begin{equation}
ds^{2}=-2dx^{+}dx^{-}-32\lambda^{2}\left[\frac{1}{4}\left((x^{1})^{2}+(x^{5})^{2}\right)^{2}+
\delta^{4}-\delta^{2}\left((x^{1})^{2}-(x^{5})^{2}\right)\right]dx^{+}dx^{+}+\sum_{i=1}^{8}dx^{i}dx^{i}
\nonumber
\end{equation}
\begin{eqnarray}
\varphi_{4}&=&\frac{\lambda}{\sqrt{2}}(x^{1}dx^{1}-x^{5}dx^{5})
(dx^{2}\wedge dx^{3}\wedge dx^{4}-dx^{2}\wedge dx^{7} \wedge
dx^{8}-dx^{4} \wedge dx^{6} \wedge dx^{7}+ \nonumber \\
&+&dx^{3}\wedge dx^{6} \wedge
dx^{8})+\frac{\lambda}{\sqrt{2}}(x^{5}dx^{1}+x^{1}dx^{5})(dx^{2}
\wedge dx^{3} \wedge dx^{8}-dx^{2} \wedge dx^{4} \wedge dx^{7}+
\nonumber \\
&+& dx^{3} \wedge dx^{4} \wedge dx^{6}-dx^{6} \wedge dx^{7} \wedge
dx^{8})
\end{eqnarray}
We see that the metric and the pure bosonic part of the light-cone
gauge lagrangian $L$ have an $SO(6)$ global symmetry, and the pure
fermionic part an $SO(8)$ symmetry. But both the bosonic-fermionic
part of $L$ and the 4-form $\varphi_{4}$ have a global symmetry
given by a group $G=SO(3)$ which rotates
%\begin{equation}
%G=\left\{(O,O), \quad O\in SO(3)\right\}\simeq SO(3),
%\label{group1}
%\end{equation}
the coordinates $2,3,4$ the same way as the coordinates $6,7,8$.
Note that this is different from the usual maximally
supersymmetric pp-wave where the global symmetry of the light-cone
gauge lagrangian is $SO(4)\times SO(4)$, or the flat space with
little group $SO(8).$

Let us return now to zero modes. Using the zero modes
anti-commutation relations of the worldsheet fermions
(\ref{zeromodes}) we obtain the Clifford algebra satisfied by the
GS fermions zero modes:
\begin{equation}
\left\{S_{0}^{a},S_{0}^{b}\right\}=\left\{\tilde{S}_{0}^{a},\tilde{S}_{0}^{b}\right\}=\frac{1}{2}A^{ab},
\quad \left\{S_{0}^{a},\tilde{S}_{0}^{b}\right\}=0 \quad
a,b=1,2,...,8 \label{cliff1}
\end{equation}
where the $8\times 8$ matrix $A$ has the form:
\begin{equation}
A=
\begin{pmatrix}%{ll}
B & 0 \\
0 & B
\end{pmatrix}
,
\end{equation}
with
\begin{equation}
B=
\begin{pmatrix}%{llll}
3 &  1 & 1 & 1 \\
1 &  3 & -1 & -1 \\
1 & -1 & 3 & -1 \\
1 & -1 & -1 & 3
\end{pmatrix}
\end{equation}
This is different than the corresponding algebra in the usual IIB
superstring ($\lambda=0$) which we would have obtained having all
worldsheet fermionic zero modes present, corresponding to
replacing $A_{ab}$ in (\ref{cliff1}) with $4\delta^{ab}$. The rank
of the matrix $A$ in (\ref{cliff1}) is only six, which means that
the Clifford algebra (\ref{cliff1}) is degenerate. To find an
irreducible representation of the Clifford algebra (\ref{cliff1})
let us consider an $SO(8)$ rotation given by $R$ which changes the
spinors $S,\tilde{S}$ into $\chi,\tilde{\chi}$:
\begin{equation}
R=
\begin{pmatrix}%{ll}
0 & C \\
-C & 0
\end{pmatrix}
, \quad \hbox{with} \quad C=\frac{1}{2}\begin{pmatrix}%{llll}
1 & -1 & -1 & -1\\
-1 & 1 & -1 & -1\\
-1 & -1 & 1 & -1\\
-1 & -1 & -1 & 1
\end{pmatrix}
\end{equation}
In this basis the fermionic zero modes satisfy the algebra:
\begin{equation}
\left\{\chi_{0}^{a},\chi_{0}^{b}\right\}=\left\{\tilde{\chi}_{0}^{a},\tilde{\chi}_{0}^{b}\right\}=2D^{ab},
\quad \left\{\chi_{0}^{a},\tilde{\chi}_{0}^{b}\right\}=0 \quad
a,b=1,2,...,8
\end{equation}
with
\begin{equation}
D=\begin{pmatrix}%{llll}
0 & & & \\
  & I_{3} & & \\
  & & 0 & \\
  & & & I_{3}
\end{pmatrix}
,
\end{equation}
where $I_{3}$ is the $3\times 3$ identity matrix. As usual we can
form fermionic oscillators:
\begin{equation}
b_{a}=\frac{\chi^{a}_{0}+i\tilde{\chi}_{0}^{a}}{\sqrt{2}}, \quad
b_{a}^{+}=\frac{\chi^{a}_{0}-i\tilde{\chi}_{0}^{a}}{\sqrt{2}}
\quad a=1,...8
\end{equation}
They satisfy the anti-commutation relations:
\begin{equation}
\left\{b_{a},b_{b}^{+}\right\}=D_{ab}, \quad
\left\{b_{a},b_{b}\right\}=\left\{b_{a}^{+},b_{b}^{+}\right\}=0
\end{equation}

We therefore obtain 6 independent fermionic oscillators, giving a
representation containing a number of \emph{64} states, \emph{32}
bosonic \emph{32} fermionic. Due to the fact that there are two
bosonic states coming from the 2d lagrangian $L_{2}$, the total
number of states of the supergravity part of superstring spectrum
is \emph{128}, half of them bosonic, half fermionic states. Under
the ``little group" $G$ defined above we can classify the
\emph{64} states coming from the Clifford algebra. Since $G$
contains elements (rotations) acting identically on both the set
of coordinates $2,3,4$ and the set $6,7,8$, we do not have
space-time fields indexed by indices belonging to different sets
(mixed indices), as we would have for a group $SO(3)\times SO(3).$
The surviving type IIB supergravity modes have the space-time
field interpretation under the group $G$ as:

\textbf{graviton}: $\quad g_{ij}(5), \quad g_{i'j'}(5)$

\textbf{complex scalar field}: $\quad \phi(2)$

\textbf{complex two-form field}: $\quad b_{ij}(6), \quad
b_{i'j'}(6)$

\textbf{four-form field}: $\quad a_{ij}(3), \quad a_{i'j'}(3),
\quad a_{1}, \quad a_{2}$

\textbf{spin 1/2 field}: $\quad \lambda_{1}(4), \quad
\lambda_{2}(4)$

\textbf{spin 3/2 field}: $\quad \psi_{i}(12), \quad
\psi_{i'}(12),$ \\
where $i,j,k=2,3,4$ and $i',j',k'=6,7,8$, and in brackets we
indicate the number of independent components. The two $\lambda's$
are complex two-component spinors of $SO(3)$. From a general
four-form field of IIB supergravity $a_{\mu\nu\alpha\beta}$
($\mu,\nu,\alpha,\beta=1,...,8$) less components can survive under
$G$. A mixed component $a_{i';ijk}$ can have non-vanishing
components from the internal symmetry of indices $i,j,k$ giving
the above antisymmetric tensor $a_{ij}$. The same is valid for
$a_{i;i'j'k'}$ giving the term $a_{i'j'}$. The other mixed
components $a_{ij;i'j'}$ give two real scalar fields
$a_{1},a_{2}$. The total number of \emph{128} states of the
supergravity part of the string spectrum is represented by two
such sets of fields in space-time.

In general the space-time interpretation of the supergravity part
of the string spectrum is that it is isomorphic with the
fluctuation modes of type IIB supergravity fields expanded near
the plane-wave type background given in (\ref{background}).
Solving analytically the equations satisfied by the above
fluctuations is impossible because they are non-linear. However,
there are in total only 128 supergravity modes in the superstring
spectrum instead of the usual 256 IIB supergravity modes. It
appears that some of the IIB supergravity modes are suppressed
when we obtain the supergravity part of the string spectrum by
taking the limit $p_{-} \to 0$, i.e UV limit of the worldsheet
theory. This situation is as if the target space is effectively
compactified on a Calabi-Yau manifold. It would be interesting to
find such a compactified string theory dual to the one we
considered above.

\section{Other backgrounds giving integrable theories on the
worldsheet}

\subsection{Superstring theories with more than one interacting
direction}

We can also consider superstring theories of the above type with
deformed cubic superpotentials but with 2,3 or 4 interacting 2d
quantum field theories. Consider the superpotential:
\begin{equation}
W=\lambda_{1}\left(\frac{1}{3}(Z^{1})^{3}-
\delta_{1}^{2}Z^{1}\right)+\lambda_{2}\left(\frac{1}{3}(Z^{2})^{3}-
\delta_{2}^{2}Z^{2}\right)
\end{equation}
The background in real coordinates is:
\begin{eqnarray*}
ds^{2}&=&-2dx^{+}dx^{-}-32\lambda_{1}^{2}\left[\frac{1}{4}\left((x^{1})^{2}+
(x^{5})^{2}\right)^{2}+
\delta_{1}^{4}-\delta_{1}^{2}\left((x^{1})^{2}-(x^{5})^{2}\right)\right]dx^{+}dx^{+} \\
&-&32\lambda_{2}^{2}\left[\frac{1}{4}\left((x^{2})^{2}+
(x^{6})^{2}\right)^{2}+
\delta_{2}^{4}-\delta_{2}^{2}\left((x^{2})^{2}-(x^{6})^{2}\right)\right]dx^{+}dx^{+}+
\sum_{i=1}^{8}dx^{i}dx^{i}
\end{eqnarray*}
\begin{eqnarray}
\varphi_{4}&=&\left(\frac{\lambda_{1}x^{1}-\lambda_{2}x^{2}}{\sqrt{2}}\right)(dx^{1}
\wedge dx^{2}+dx^{5} \wedge dx^{6})(dx^{3} \wedge dx^{4}-dx^{7}
\wedge dx^{8}) \nonumber \\
&-&\left(\frac{\lambda_{1}x^{5}-\lambda_{2}x^{6}}{\sqrt{2}}\right)(dx^{5}
\wedge dx^{2}-dx^{1} \wedge dx^{6})(dx^{3} \wedge dx^{4}-dx^{7}
\wedge dx^{8}) \nonumber \\
&+&\left(\frac{\lambda_{1}x^{5}-\lambda_{2}x^{6}}{\sqrt{2}}\right)(dx^{1}
\wedge dx^{2}+dx^{5} \wedge dx^{6})(dx^{7} \wedge dx^{4}+dx^{3}
\wedge dx^{8}) \nonumber \\
&+&\left(\frac{\lambda_{1}x^{1}-\lambda_{2}x^{2}}{\sqrt{2}}\right)(dx^{5}
\wedge dx^{2}-dx^{1} \wedge dx^{6})(dx^{7} \wedge dx^{4}+dx^{3}
\wedge dx^{8})
\end{eqnarray}
The global symmetry of this background is given by a group
$G_{1}=SO(2)$ which rotates the coordinates $3,4$ the same way as
the coordinates $7,8$.

On the worldsheet the above background gives two interacting 2d
theories in $Z^{1},Z^{2}$ and free theories in $Z^{3},Z^{4}.$ For
each of the interacting theories the exact energy eigenvalues can
be computed and represented as in Figure 1. Therefore the full
energy eigenvalue configuration of the light-cone gauge string
hamiltonian can be obtained as depending on two dimensionless
parameters $\frac{4}{3}\lambda_{1}\delta_{1}^{3}|p_{-}|$ and
$\frac{4}{3}\lambda_{2}\delta_{2}^{3}|p_{-}|.$ As in the previous
sections the ground state of the light-cone gauge string
hamiltonian contains no excited states from the interacting parts
and no oscillatory states from the free parts. Now there are fewer
fermionic zero modes than in the previous section but there are
four bosonic states. By proceeding as in section 4 we can find the
``massless" string states. Due to the degeneracy of the zero-modes
GS fermions we have \emph{16} states in space-time, half of them
bosonic, half fermionic states. The surviving states of IIB
supergravity can be classified in space-time under the group
$G_{1}$ as:

\textbf{graviton}: $\quad g_{ij}(2), \quad g_{i'j'}(2)$

\textbf{antisymmetric tensor}: $\quad b_{ij}(1), \quad
b_{i'j'}(1)$

\textbf{complex scalar field}: $\quad \phi (2)$

\textbf{spin 1/2 field}: $\quad \lambda_{1}(2), \quad
\lambda_{2}(2)$

\textbf{spin 3/2 field}: $\quad \psi_{1}(2), \quad
\psi_{2}(2),$ \\
where $i,j=3,4$ and $i',j'=7,8$, and the two $\lambda's$ are
complex one-component Weyl spinors of $SO(2).$ Since the bosonic
part of $H$ has a fourfold degeneracy we have a total of \emph{64}
``massless" string states represented by four sets of fields each
containing the above fields.

We may take now the superpotential of the form:
\begin{equation}
W= \sum_{i=1}^3 \lambda_{i}\left(\frac{1}{3}(Z^{i})^{3}-
\delta_{i}^{2}Z^{i}\right)
\end{equation}
By similar procedures as before we see that there are 4 states
from the Clifford algebra of fermionic zero modes. In this case
the global symmetry of the background is $SO(1)\times SO(1)$.
Under this group we can have only a complex scalar field $\phi$
and two real one-component $SO(1)$ spinors
$\lambda_{1},\lambda_{2}$. Since the bosonic degeneracy is now 8,
the total number of ``massless" string states in space-time is
\emph{32} (half bosonic, half fermionic states). There are eight
sets of fields each consisting in a complex scalar field and two
spinors.

Lastly we can consider the superpotential:
\begin{equation}
W= \sum_{i=1}^4 \lambda_{i}\left(\frac{1}{3}(Z^{i})^{3}-
\delta_{i}^{2}Z^{i}\right)
\end{equation}
Now there are no free parts in the light-cone gauge hamiltonian.
Also, for the ground state there is no fermionic degeneracy, and
therefore there is a total of \emph{16} bosonic ``massless" string
states which correspond to \emph{16} real scalar fields in
space-time.

As in section 4 in all the above cases the supergravity part of
superstring spectrum, obtained by taking the limit $p_{-} \to 0$,
appears not to be in one-to-one correspondence to type IIB
supergravity modes, as some of these modes are suppressed.
However, in the limit $\alpha' \to 0$ we obtain the full IIB
supergravity spectrum in all the above cases.

\subsection{Superstring theory on backgrounds giving a ${\cal
N}=(2,2)$ supersymmetric sine-Gordon}

One can also consider the superstring theory obtained by choosing
the superpotential $W=\frac{M}{\omega^{2}} \cos (\omega Z^{1}).$
The background in this case is given by:
\begin{equation}
ds^{2}=-2dx^{+}dx^{-}-32\left|\frac{M}{\omega} \sin (\omega
z^{1})\right|^{2}dx^{+}dx^{+}+2dz^{i}d\bar{z}^{\bar{i}} \nonumber
\end{equation}
\begin{equation}
F_{5}=dx^{+}\wedge \varphi_{4}, \quad \varphi_{4}=-M\cos(\omega
z^{1})dz^{1}\wedge d\bar{z}^{\bar{2}}\wedge
d\bar{z}^{\bar{3}}\wedge d\bar{z}^{\bar{4}}+c.c.
\end{equation}
Again we can rewrite the worldsheet action in terms of
dimensionless fields by rescaling the fields and parameters
($Z^{1}\equiv Z$): $Z\rightarrow \sqrt{2\pi\alpha'}Z$;
$\omega\rightarrow \frac{\beta}{\sqrt{2\pi\alpha'}}.$ The
interacting part of light-cone gauge action is given by:
\begin{equation}
S_{2}=\frac{1}{2}\int d\tau \int_{0}^{2\pi\alpha'|p_{-}|}d \sigma
L_{2} \nonumber
\end{equation}
\begin{eqnarray}
L_{2}&=&\partial_{\tau}Z
\partial_{\tau}\bar{Z}-
\partial_{\sigma}Z \partial_{\sigma}\bar{Z}-
\frac{M^{2}}{4\beta^{2}}|\sin \beta
Z|^{2}+i\bar{\psi}_{+}(\partial_{\tau}-\partial_{\sigma})\psi_{+}+
i\bar{\psi}_{-}(\partial_{\tau}+\partial_{\sigma})\psi_{-}+ \nonumber \\
&-&\frac{i}{2}M\cos(\beta
Z)\psi_{+}\psi_{-}-\frac{i}{2}M\cos(\beta\bar{Z})\bar{\psi}_{-}\bar{\psi_{+}},
\end{eqnarray}
In the weak coupling limit when $\beta\rightarrow 0$ or in string
scale $\omega\rightarrow 0$, one can keep the first terms in the
expansion of the superpotential and obtain $W=\frac{MZ^{2}}{2}.$
For this superpotential the equations of motion of the light-cone
gauge lagrangian can be solved analytically. In terms of real
coordinates and GS spinors $S,\tilde{S}$, the light-cone gauge
lagrangian density $L$ contains two free massive scalars
($X^{1},X^{5}$), six free massless scalars while all the fermions
are massive and free. This can be solved exactly in a similar way
as in the case of the usual pp-wave background. Thus, in the limit
$\beta\rightarrow 0$, the supergravity modes of the string
spectrum are the IIB supergravity modes, as expected for weak
coupling. Further, as in the case of maximally supersymmetric
pp-wave, the IIB supergravity spectrum is in one-to-one
correspondence with the IIB supergravity fluctuation modes around
the background under consideration.

Let us consider now the supergravity limit $\alpha' \to 0$ while
everything else is fixed ($M,\omega, p_{-}$). In this limit the
worldsheet theory reduces to a free massive theory (like in the
case of $z^3-z$ superpotential) with mass $M$, as $\beta \to 0$ in
the $\alpha' \to 0$ limit. As we already mentioned at the
beginning of section 4, from the worldsheet point of view we have,
in the limit $\beta\rightarrow 0$, an infinite number of breather
states with masses $M, 2M, 3M,\dots.$ As in the case of the
superpotential $z^3-z$ we obtain the full IIB supergravity
spectrum in the limit $\alpha' \to 0$. The supergravity part of
the string spectrum has the hamiltonian:
\begin{equation}
H_{0}=\frac{1}{2|p_{-}|}\sum_{\alpha}^{}p^{\alpha}p^{\alpha}+fM,
\label{supgrav1}
\end{equation}
where $f$ is an integer number depending on the level of
excitation of the harmonic oscillators, and $\alpha$ runs over the
free coordinates $2,3,4,6,7,8.$ Let us now consider solving the
scalar fluctuation $\Phi$ belonging to the massless supergravity
multiplet. The curved-space Klein-Gordon equation satisfied by
this fluctuation $\Box \Phi =0$ has been considered in
\cite{russo1}. It has been shown that the light-cone gauge
hamiltonian corresponding to the supergravity part of superstring
spectrum is given by:
\begin{equation}
H_{0}=\frac{1}{2|p_{-}|}\sum_{\alpha}^{}
p^{\alpha}p^{\alpha}+\frac{1}{2|p_{-}|}(a_{r}+e_{n}),
\label{supgrav}
\end{equation}
where the discrete set of values $a_{r}$ and $e_{n}$ depend on
$M,\omega$, and can be determined by solving two differential
equations having solutions expressed in terms of Mathieu
functions. For arbitrary values of $M,\omega$ this hamiltonian
does not match the ``massless" string states hamiltonian
(\ref{supgrav1}). Therefore it seems that the IIB supergravity
spectrum is not in one-to-one correspondence to the IIB
supergravity fluctuation modes around the background under
consideration unless $\omega \to 0$, which reduces to the above
disscused weak coupling limit $\beta \to 0$.

Let us consider what happens when we take the limit $p_{-} \to 0$
while keeping everything else fixed ($\beta \neq 0$). For
$\beta^{2}< 8\pi$ the spectrum of the worldsheet theory consists
of kinks, antikinks and a finite number of breathers. We, however,
do not have the breather state energies. As we mentioned in
section 4 computing these energies as functions of $\beta$ and the
worldsheet size $R$ would be interesting. The regime where we can
analyze the behavior of the energy levels in the limit $p_{-}\to
0$ is for $\beta^{2}\ge 8\pi.$ In this case, when considered on
two dimensional non-compact space, the supersymmetric sine-Gordon
theory has a spectrum consisting of only kinks and anti-kinks with
masses $m=\frac{2M}{\beta^{2}}$, and no stable breathers
(kink-antikink bound states). When the direction $x^{1}$ is
compact, with identification $x^{1}\cong x^{1}+2\pi/\omega.$, the
kinks and anti-kinks interpolate between the two supersymmetric
vacua at $x^{1}=0,x^{5}=0$ and $x^{1}=\pi/\beta,x^{5}=0$, or in
string scale at $x^{1}=0,x^{5}=0$ and $x^{1}=\pi/\omega,x^{5}=0.$
The dimensionless parameter is $mR=\frac{2M|p_{-}|}{\omega^{2}}.$
We need to find the spectrum of this theory on a cylinder of
circumference $R=2\pi\alpha'|p_{-}|.$ The excited levels of the
interacting part $H_{2}$ of the light-cone gauge hamiltonian can
be computed exactly by the method in \cite{fendley4}. While we do
not have explicit results for arbitrary $\beta$, we have checked
the values $\beta^2=8\pi$ and $16\pi$, where the analysis is
simplest. We have found that the dependence of the excited levels
on $mR$ looks similar to those presented in figure 1. As in
section 3 they all decouple in the point particle limit $p_{-}\to
0.$ By proceeding in a similar manner as in section 4 we see that
there are \emph{128} modes corresponding to the supergravity part
of the superstring spectrum. This again implies that the full IIB
supergravity spectrum is not obtained, as some of these modes as
missing.

\section{Conclusions}

We analyzed the spectrum of string theories on a class of pp-wave
backgrounds which result in ${\cal N}=(2,2)$ supersymmetric
interacting theories on the worldsheet given by the superpotential
$z^{3}-z.$ The energy eigenvalue configuration of the light-cone
gauge string hamiltonian can be determined exactly. Knowing these
energies will hopefully be useful in finding the full string
spectrum. An interesting limit which may be connected to strings
on discrete target spaces and perhaps to non-critical string
theories, is obtained when the kink mass $m\rightarrow\infty.$ In
this limit (or strings with very large $p_{-}$) the string, in its
excited states, does not see two of the eight transverse
directions. We found the supergravity modes of the string spectrum
obtained by taking the $\alpha' \to 0$ and $p_{-} \to 0$ limit. In
the former case we found the full type IIB supergravity spectrum,
as expected. However, in the case of the second limit, $p_{-} \to
0$, which is the UV limit on the worldsheet, we do not obtain the
full IIB supergravity spectrum: as we have detailed, some of the
modes are suppressed. In this case we analyzed the ``massless"
string modes and their space-time field interpretation under the
global symmetry of the background given by the group $G$. The
absence of some modes of IIB supergravity spectrum, obtained as
$p_{-} \to 0$, seems also to be true when the worldsheet theory is
the supersymmetric sine-Gordon model, at least in the limit when
$\beta^{2}\ge 8\pi$, or strong enough coupling $\beta.$

\bigskip\bigskip
We are grateful to K.~Intriligator for bringing this problem to
our attention, and for useful conversations. We are also grateful
to J.~Maldacena for his comments, in particular on disappearing
supergravity modes. This work was supported by a Research
Corporation Research Innovation award, as well as the DOE under
grant DEFG02-97ER41027 and the NSF under grant DMR-0104799.

\end{document}